\documentclass{article}

\usepackage{PRIMEarxiv}

\usepackage[utf8]{inputenc} 
\usepackage[T1]{fontenc}    
\usepackage{hyperref}       
\usepackage{multirow}
\usepackage{url}            
\usepackage{booktabs}       
\usepackage{amsfonts}       
\usepackage{nicefrac}       
\usepackage{microtype}      
\usepackage{lipsum}
\usepackage{authblk}         
\usepackage{orcidlink}       
\usepackage{fancyhdr}       
\usepackage{graphicx}       
\graphicspath{{media/}}     

\title{Deep Ensemble approach for Enhancing Brain Tumor Segmentation in Resource-Limited Settings}

\author[1,2]{Jeremiah Fadugba\orcidlink{0009-0008-1526-4195}}
\author[3]{Isabel Lieberman\orcidlink{0000-0002-7228-8547}}
\author[4]{Olabode Ajayi\orcidlink{0000-0002-9496-4227}}
\author[10]{Mansour Osman}  
\author[3]{Solomon Oluwole Akinola\orcidlink{0000-0002-4818-6348}}
\author[3]{Tinashe Mustvangwa}
\author[9]{Dong Zhang}
\author[3,7,8,9,6]{Udunna C Anazondo}  
\author[7,8]{Raymond Confidence}

\affil[1]{University of Ibadan, Ibadan, Nigeria}
\affil[2]{African Institute for Mathematical Sciences, Kigali, Rwanda}
\affil[3]{University of Cape Town, Cape Town, South Africa}
\affil[4]{South African National Bioinformatics Institute (SANBI), University of Western Cape, Cape Town, South Africa}
\affil[5]{Institute for Intelligent Systems, University of Johannesburg, Johannesburg, South Africa}
\affil[10]{Carnegie Mellon University, Kigali, Rwanda}
\affil[6]{Medical Artificial Intelligence Laboratory (MAI Lab), Lagos, Nigeria}
\affil[7]{Lawson Health Research Institute, London, Ontario, Canada}
\affil[8]{Department of Electrical and Computer Engineering, University of British Columbia, Vancouver, Canada}
\affil[9]{Montreal Neurological Institute, McGill University, Montréal, Canada}

\begin{document}
\maketitle

\begin{abstract}
Segmentation of brain tumors is a critical step in treatment planning, yet manual segmentation is both time-consuming and subjective, relying heavily on the expertise of radiologists. In Sub-Saharan Africa, this challenge is magnified by overburdened medical systems and limited access to advanced imaging modalities and expert radiologists. Automating brain tumor segmentation using deep learning offers a promising solution. Convolutional Neural Networks (CNNs), especially the U-Net architecture, have shown significant potential. However, a major challenge remains: achieving generalizability across different datasets. This study addresses this gap by developing a deep learning ensemble that integrates UNet3D, V-Net, and MSA-VNet models for the semantic segmentation of gliomas. By initially training on the BraTS-GLI dataset and fine-tuning with the BraTS-SSA dataset, we enhance model performance. Our ensemble approach significantly outperforms individual models, achieving DICE scores of 0.8358 for Tumor Core, 0.8521 for Whole Tumor, and 0.8167 for Enhancing Tumor. These results underscore the potential of ensemble methods in improving the accuracy and reliability of automated brain tumor segmentation, particularly in resource-limited settings.
\end{abstract}


\section{Introduction}

Brain tumors are typically abnormal growth in brain cells, and they represent one of the most severe and life-threatening conditions, necessitating precise medical imaging for effective diagnosis and treatment \cite{Havaei2017-ru}. In clinical practice, the delineation of tumor is usually done manually by experienced radiologists studying each scan of the medical image. This process can be time-consuming and subjective to radiologists, leading to significant intra- and inter-rater variability \cite{Isin2016-xd}. The manual segmentation of brain tumor arguably provides a more accurate segmentation, but due to the large variability in shape, size, location of lesions and visually inspecting more than one image modalities, this becomes impractical for more extensive studies \cite{magadza_deep_2021}. Accurate tumor segmentation is crucial for determining the extent of the disease and planning surgical interventions, radiotherapy, and ongoing patient monitoring \cite{sun2019drrnet}. However, achieving reliable segmentation remains a significant challenge, especially in resource-limited settings like Sub-Saharan Africa (SSA) \cite{Kanmounye2022-lp}.

In SSA, the burden of brain tumors is compounded by limited access to advanced medical imaging technologies and expert radiologists \cite{adewole2023status}. This shortage significantly hampers early diagnosis and appropriate treatment, leading to poorer prognoses compared to high-income countries \cite{anazodo2022ai}\cite{Kanmounye2022-lp}. The manual segmentation challenge the global north faces is further exacerbated in SSA due to an already overburdened system. Conventional automated techniques, such as thresholding and region-growing, have limitations in handling the complex and heterogeneous nature of brain tumors. These methods often fail to capture subtle tumor boundaries and variations in tumor appearance, necessitating the development of more advanced and reliable segmentation techniques.

Recently, many efforts have been made to automate the segmentation of brain tumor from multi-modal Magnetic Resonance Imaging (MRI). Most of these methods are based on deep learning, particularly convolutional neural networks (CNNs) \cite{abd2024automatic}. The state-of-the-art methods for brain tumor segmentation are based on the UNet architecture \cite{ronneberger2015u,abd2024automatic}, which is the most famous architecture in medical imaging \cite{somasundaram_current_2019} \cite{magadza_deep_2021}, have demonstrated exceptional performance in various image recognition tasks. Similarly, in recent times, the segment anything model (SAM) \cite{huang2024segment,barakat2023towards} has shown remarkable performance in brain tumor segmentation. These models learn hierarchical features from the data, enabling them to generalize well across different imaging conditions and patient populations. 

Despite the progress made with deep learning approaches, several gaps remain in the current literature. Existing models often struggle with generalizability across diverse datasets, particularly in SSA where imaging conditions and patient demographics differ significantly from those in high-income countries \cite{adewole2023brain}. Furthermore, there is a need for models that are not only accurate but also robust and efficient, capable of operating in resource-constrained environments with limited computational power \cite{futrega_optimized_2021} \cite{barakat_towards_2023}. Addressing these gaps requires novel methodologies that can enhance the performance and applicability of deep learning models in brain tumor segmentation. This research aims to contribute to this endeavour by developing and validating a deep learning-based segmentation method tailored to the unique challenges of the SSA dataset, ultimately striving to improve clinical outcomes in this underserved region.

Our findings indicate that:
\begin{enumerate}
    \item The deep learning ensemble approach, which incorporates several models, outperforms individual models, resulting in better assessment metrics.

    \item The BraTS-Africa dataset segmentation indicates DICE scores of 0.8167, 0.8358, and 0.8521 for the Enhancing Tumor, Tumor Core and Whole Tumor scores, respectively.
    
\end{enumerate}

\section{Related Works}
Brain tumor research in low- and middle-income nations has been hindered by the limited availability of MRI scanners \cite{murali2023bringing}. Glioma mortality rates are notably high in Sub-Saharan Africa, highlighting the urgent need for improved access to advanced imaging and multidisciplinary treatment. Manually segmenting brain tumors from MRI images for tumor detection is a demanding and time-consuming task that is prone to variation between different observers \cite{razzak2018efficient,sun2019drrnet}. This variation can greatly impact the accuracy and consistency of the segmentation results. Since a single brain MRI scan consists of multiple slices that collectively form a 3D anatomical view, the manual segmentation of brain tumor MR images becomes a complex procedure \cite{sun2019drrnet}. Furthermore, computer-aided diagnosis through machine learning can significantly improve tumor diagnostic accuracy, early detection, classification, and prognosis of patient survival rates.

Recent studies have shown that deep learning CNN-based auto-segmentation models can significantly improve efficiency and effectiveness. Conventional deep learning methods, such as CNN, require substantial amounts of labelled data for optimal learning, which presents challenges in the medical domain \cite{razzak2018efficient}. Deep-learning CNNs have proven to be highly valuable in accurately segmenting various structures during the treatment planning phase \cite{liang2020emerging,ibragimov2017segmentation,lustberg2018clinical,jackson2018deep,hu2016automatic,ibragimov2017combining}.
Again, ensemble techniques for brain tumor segmentation further enhance whole tumor, tumor core, and enhancing tumor on the test dataset, respectively \cite{khan2023hybrid,koirala2023automated}.

\section{Methodology}

\subsection{Dataset Description}

The BraTS-Africa data set \cite{adewole2023brain} includes imaging data from a significant number of African patients. For training, 60 patient scans are made available. Each patient's scans include image volumes of T1-weighted (T1), post gadolinium (Gd) contrast T1-weighted (T1Gd), T2-weighted (T2), and T2-Fluid Attenuated Inversion Recovery (T2-FLAIR).

The dataset was collected using standardized imaging protocols, with adjustments made to suit the African population's specific needs. The scans were performed using advanced MRI machines, ensuring high-quality data. The precise models and manufacturers of the MRI machines used are not detailed.


The BraTS-Africa dataset's tumor annotation protocol ensures consistent ground truth labels using the BraTS standard (cite). Initial automated segmentations were generated using a nnU-Net model and refined manually by trained radiologists with varying experience. The refined segmentations were reviewed iteratively by senior board-certified radiologists using the ITK-SNAP software until deemed acceptable for public release and the challenge. The entire process and segmentation methods are available on the Federated Tumor Segmentation (FeTS) platform.


The dataset can be assessed via the synapse platform\footnote{\url{https://www.synapse.org/\#!Synapse:syn51156910/wiki/622556}}. The BraTS-Africa dataset was gathered with the support of the Consortium for Advancement of MRI Education and Research in Africa (CAMERA)\footnote{\url{https://www.cameramriafrica.org/}} and funding from the Lacuna Fund in Health Equity\footnote{\url{https://lacunafund.org/}}.

\subsection{Pre-processing} 

In addition to the standard pre-processing pipeline from the BraTS Challenge organizers \cite{adewole2023brain}, we made further pre-processing of the scans following \cite{ren_optimization_2024}. 

\subsubsection*{Z-Score Normalization}

Z-score normalization was employed to manage the varying intensity distributions observed across the dataset, as referenced in \cite{luu2021extendingnnunetbraintumor}. This technique involves calculating the mean and standard deviation of voxel intensities, then transforming each voxel by subtracting the mean and dividing by the standard deviation. By standardizing the intensity values of each voxel, Z-score normalization facilitates better direct comparisons between different patient datasets and effectively removes outliers \cite{ren_optimization_2024}.



\subsubsection*{Rescaling Voxel Intensitities}
Segmentation efficacy hinges on the ability to discern critical features and structures. To enhance the visibility of these features, voxel intensities were scaled. We calculated the voxel intensity percentiles and defined the 2nd and 98th percentiles as the range for intensity stretching, extending this range to cover the entire intensity spectrum \cite{ren_optimization_2024}. This method increases contrast and provides greater clarity of subtle features within the data.


\subsection{Network Architecture}

\subsubsection{UNet3D} is an extension of the 2D U-Net architecture, specifically designed for volumetric medical image segmentation. It leverages 3D convolutions to capture spatial context across all three dimensions, thereby improving the accuracy of segmentation for complex structures within medical images. The architecture consists of an encoder-decoder structure with skip connections that allow for the preservation of fine-grained details while progressively capturing higher-level features. This makes UNet3D particularly effective for tasks such as brain tumor segmentation, where understanding the spatial relationships within the volume is crucial.

\begin{figure}[ht]
    \centering
    \includegraphics[width=0.6\linewidth]{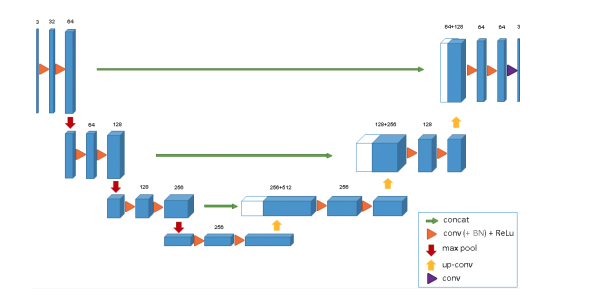}
    \caption{Unet3D architecture \cite{unetlearningdense}}
    \label{fig:unet}
\end{figure}

\subsubsection{V-Net}
\begin{figure}[ht]
    \centering
    \includegraphics[width=0.6\linewidth]{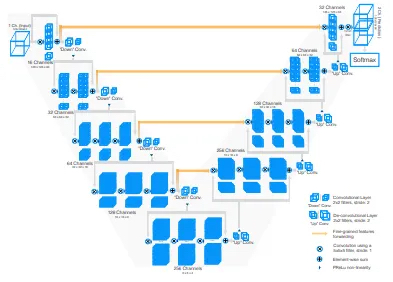}
    \caption{V-Net architecture \cite{milletari2016vnetfullyconvolutionalneural}}
    \label{fig:vnet}
\end{figure}

The V-Net architecture (see figure \ref{fig:vnet} ) is a similar network to UNet3D and its also designed for volumetric medical image segmentation \cite{milletari2016vnetfullyconvolutionalneural}. It features an encoder-decoder structure resembling a 'V' shape, hence its name. With a major similarity to Unet3D, V-Net differs in the kernel sizes by using a volumetric kernels having size $5 \times 5 \times 5$ voxels withing its residual blocks to enhance feature extraction. Skip connections in V-Net sums the input features with the output of the convolutional layers, allowing the network to combine low-level spatial information with high-level semantic information. In this work, We modify the original V-Net architecture \cite{milletari2016vnetfullyconvolutionalneural} by first increasing the number of feature maps from 16 to 32 in contrast to the original implementation.

\subsubsection{Multi-Scale Attention VNet(MSA-VNet:}We implemented MSA-VNet 
We implemented MSA-VNet (Multi-Scale Attention V-Net)\cite{xu_msa-vnet_2022}, an advanced 3D convolutional neural network for volumetric medical image segmentation that builds upon V-Net principles. This architecture incorporates multi-scale attention mechanisms to enhance feature representation and segmentation accuracy. It consists of an encoder-decoder structure, where the encoder compresses input through convolutional and pooling layers, capturing multi-scale features. A central convolutional block refines deep features before the decoder path, which uses transposed convolutions for up-sampling and integrates encoder information via skip connections. Multi-scale attention blocks at each skip connection allow the model to focus on relevant features from both paths, improving its ability to distinguish subtle input differences. The network concludes with a final convolutional layer outputting the segmented map, resulting in a powerful tool for precise medical image segmentation.

\begin{figure}
    \centering
    \includegraphics[width=0.75\linewidth]{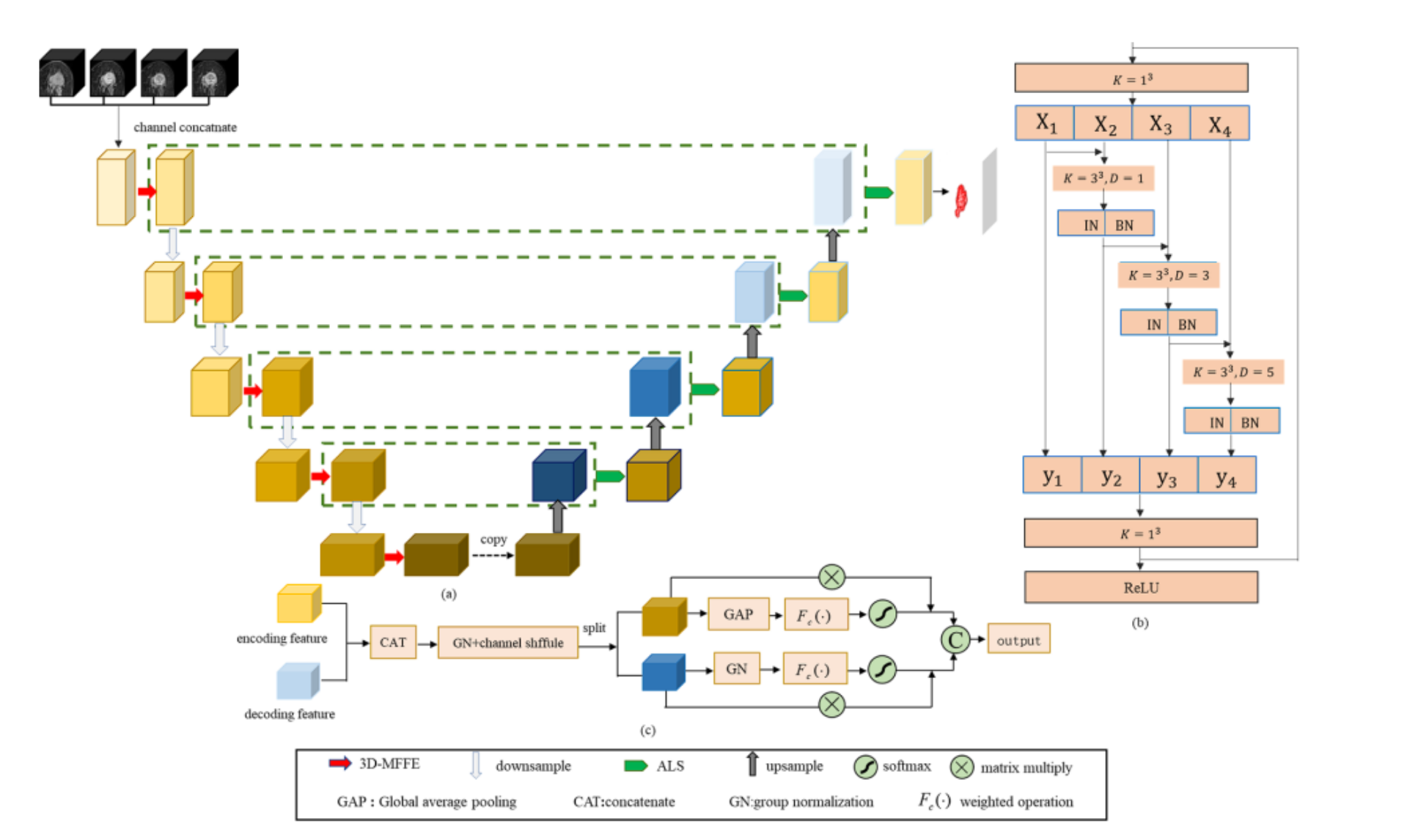}
    \caption{Multi-Scale Attention-VNet architecture}
    \label{fig:msavnet}
\end{figure}

\subsection{Deep Ensemble Learning}
Deep Ensemble learning is a common technique in machine learning and in particular, brain tumor segmentation where multiple deep learning models are trained and combined to make a single prediction \cite{isin_review_2016}\cite{jiang_brain_tumor_ensemble}\cite{ren_optimization_2024}

In this work, our approach combines the network: of UNet3D \cite{unetlearningdense}, VNet \cite{milletari2016vnetfullyconvolutionalneural}, and MSA-VNet \cite{xu_msa-vnet_2022} to achieve superior segmentation performance. As commonly used in Brain Tumor Segmenration, the final results were obtained by using the Simultaneous Truth and Performance Level Estimation (STAPLE) algorithm \cite{Warfield2004-kd}. The STAPLE method operates by first estimating both the true segmentation and the performance of each model within an ensemble. It then merges the outputs of these models to produce a final segmentation, achieving a higher accuracy compared to any single model's prediction. By integrating the predictions from these diverse models, the ensemble method reduces individual model biases and variances, leading to more robust and accurate segmentation outcomes.

\subsection{Post-Processing}
Post-processing is a crucial step in glioma segmentation that occurs after the initial tumor segmentation. This step involves refining the initial results to enhance accuracy and minimize errors or inconsistencies in the final predictions. Techniques such as morphological operations, region growing, and level set evolution are often employed. Post-processing aims to achieve the most accurate and reliable final segmentation possible.

Following the post-processing procedures in \cite{ren_optimization_2024}, all connected components of ET (enhancing tumor) voxels and remove those with a volume of 50 voxels or less, relabeling these small components as 0. Similarly, the TC (tumor core) voxels, which include both ET and NCR (non-enhancing core) voxels, to ensure that the removal of ET voxels has not created any holes in this region.

\section{Experiments and Results}

\subsection{Training Details}
All model training and development were carried out using PyTorch, a deep-learning framework. The experimental setup includes four NVIDIA GeForce RTX 2080 Ti GPUs, PyTorch version 1.8.0, and Python version 3.6.5. The models were trained using the Dice Loss function available on MONAI for 40 epochs with a batch size of 4. We employed the AdamW optimizer, incorporating a weight decay factor of $1 \times 10^{-5}$, an initial learning rate set to  $6 \times 10^{-5}$, and a cosine annealing strategy for learning rate scheduling. For finetuning the same hyperparameters were used however, for cross validation experiments, we reduce the epochs to 30 for each folds.

\subsection{Model Results}
We divided the official BraTS-GLI training dataset into a 60:20:20 split, holding $20\%$ of the dataset for validation and $20\%$ for testing. After every epoch, we iterate through this validation hold, calculating the Dice score between model prediction and the ground truth for the regions of ET, TC, and WT and then averaging these scores. 

For fine-tuning on the BraTS-SSA dataset we use the same pre-processing technique used for the BraTA-GLI dataset and the training hyper-parameters remain the same. However, we freeze the decoder for UNet3D to avoid over-fitting since it has more parameters than other models while for V-Net and MSA-VNEt, we continue training on the BraTS-SSA dataset.

The results in Table ~\ref{tab:results_with_postprocessing} shows the DSC and the Hausdorff distance (95\%) (HD95) values for a hold out test set of 12 cases from the BraTS-SSA training set. The post processing follows as discussed earlier where any small connected components were removed and the labels set to 0. The use of this post-processing technique saw just a bit of an increase in performance for each of the model (see Table ~\ref{tab:results_with_postprocessing}).

\begin{table}[ht]
    \setlength{\tabcolsep}{6pt}
    \centering
    \caption{Dice and HD95 scores for TC, WT, and ET for each models on the hold-out test cases with and without post-processing.}
    \begin{tabular}{lccccccccccc}
        \toprule
        \multirow{2}{*}{Model} & \multirow{2}{*}{Post-Processing} & \multicolumn{3}{c}{DSC $\uparrow$} & \multicolumn{3}{c}{HD95 $\downarrow$} \\
        \cmidrule(lr){3-5} \cmidrule(lr){6-8}
        & & ET & TC & WT & ET & TC & WT \\
        \midrule
        UNet & Without      & 0.7337 & 0.6456 & 0.8182 & 19.85 & 20.36 & 14.76  \\
                & With      & 0.7421 & 0.6532 & 0.8234 & 20.11 & 21.77 & 15.84 \\
        V-Net & Without     & 0.7374 & 0.7651 & 0.8041 & 9.13  & 10.14 & 17.62 \\
                & With      & 0.7381 & 0.7588 & 0.8132 & 9.56  & 11.52 & 17.34 \\
        MSA-VNet & Without  & 0.5667 & 0.5856 & 0.7483 & 11.91 & 14.22 & 21.28 \\
                & With      & 0.6430 & 0.5875 & 0.7325 & 10.89 & 14.96 & 20.32 \\
        \bottomrule
    \end{tabular}
    \label{tab:results_with_postprocessing}
\end{table}

\subsection{Online Evaluation}
Table ~\ref{tab:results:val} summarizes the results of each model on the BraTS-SSA validation set computed on the Synapse Platform\footnote{\url{https://www.synapse.org}}. The Dice score and Hausdorff distance 95\% (HD95) were computed via the platform. Our results for the three evaluated tumor sub-regions ET, WT and TC regions with the post-processing technique shows that the ensemble model gained substantially for the WT region and not so much improvement in the other regions.

\begin{table}[ht]
    \setlength{\tabcolsep}{12pt}
    \centering
    \caption{Results of each model on the BraTS 2024 SSA Validation set.  The average of DSC and HD95 scores are computed via the Synapse online platform}
     \begin{tabular}{lccccccc}
        \toprule
        Model & \multicolumn{3}{c}{DSC $\uparrow$} & \multicolumn{3}{c}{HD95 $\downarrow$}  \\ 
        & ET & TC & WT  & ET & TC & WT \\
        \midrule
        UNet     & 0.7679 & 0.7695 & 0.8203 & 22.85 & 24.21 & 11.93  \\
        V-Net    & 0.8021 & 0.8113 & 0.8348  & 16.76 & 19.38 & 19.278 \\
        MSA-VNet & 0.7960 & 0.8085  & 0.7312 & 29.05 & 31.59 & 37.55 \\
        Ensemble & 0.8167 & 0.8358 & 0.8521  & 16.86 & 18.65 & 13.44\\
        \bottomrule
    \end{tabular}
    \label{tab:results:val}
\end{table}

The performance of UNet3D, V-Net, and MSA-VNet models was evaluated using 5-fold cross-validation on the BraTS 2024 SSA Validation set. The results are presented in Table ~\ref{tab:results:CV}, showing the mean Dice scores for Enhancing Tumor (ET), Tumor Core (TC), and Whole Tumor (WT) across all folds.

VNet demonstrated superior performance across all metrics, achieving the highest mean Dice scores of 0.8365, 0.8657, and 0.8490 for ET, TC, and WT, respectively. MSA-VNet showed improved performance over the baseline UNet3D model, with mean Dice scores of 0.8066, 0.8275, and 0.8141 for ET, TC, and WT. The UNet3D model, while still performing reasonably well, had the lowest scores among the three.

Notably, all models performed well on the Whole Tumor segmentation task, followed by the Tumor Core, with the Enhancing Tumor presenting the most challenging task as evidenced by the lower Dice scores across all models. The consistent superior performance of VNet across all tumor regions suggests that its multi-scale attention mechanism effectively enhances the model's ability to segment complex tumor structures in brain MRI scans.

These results underscore the effectiveness of the proposed MSA-VNet architecture in improving segmentation accuracy for brain tumor segmentation, particularly on the Sub-Saharan Africa dataset. 

\begin{table}[ht]
    \setlength{\tabcolsep}{12pt}
    \centering
    \caption{Results of 5 Fold cross-validation of each model on the BraTS 2024 SSA Validation set. The average of DSC and HD95 scores are computed via the Synapse online platform}
    \begin{tabular}{lccccccc}
        \toprule
        Model & \multicolumn{3}{c}{DSC $\uparrow$} & \multicolumn{3}{c}{HD95 $\downarrow$}  \\
        & ET & TC & WT  & ET & TC & WT \\
        \midrule
        UNet     & 0.7894 & 0.7941  & 0.8326 & 21.13 & 22.73 & 12.05 \\
        V-Net    & 0.8365 & 0.8567  & 0.8490 & 15.23 & 17.38 & 21.32 \\
        MSA-VNet & 0.8066 & 0.8275  & 0.8141 & 26.12 & 27.94 & 22.06 \\
        \bottomrule
    \end{tabular} 
    \label{tab:results:CV}
\end{table}

\section{Conclusion}
This paper presented our contribution to the brain tumor segmentation task on data from low resource settings with significantly different resolutions. This is a unique opportunity provided by the BraTS-Africa Challenge. This also is essential in developing and evaluating deep learning methods for brain tumors management in resource-limited settings. With the numerous challenges posed by the peculiarity of the BraTS-Africa dataset; such as the image quality and the number of cases, better performance was achieved with individual deep learning model. Moreover our study demonstrates the effectiveness of ensemble models.  It is also noteworthy to mention that UNet3D which served as a baseline competes favourably with other models. However, our investigation into a different architecture; V-Net showed a 4.5\%, 5.4\% and 1.8\% increase in Dice score for ET, TC, and WT, respectively. We considered how attention mechanism can help push the performance using MSA-VNet but this architecture did not give any improvement to the baseline UNEt3D and the V-Net model. Overall, we see that the ensemble approach gave a 6.4 \%, 8.6\% and 3.9\% performance increase in Dice score for ET, TC, and WT, respectively.

\section*{Acknowledgments}
The authors would like to thank the following instructors of the Sprint Al Training for African Medical Imaging Knowledge Translation (SPARK) Academy 2023 summer school on deep learning in medical imaging for providing insightful background knowledge on brain tumors that informed the research presented here. The authors would also like to thank Linshan Liu for administrative assistance in supporting the SPARK Academy training and capacity building activities which the authors immensely benefited from. The authors acknowledge the computational infrastructure support from the Digital Research Alliance of Canada (The Alliance) and knowledge translation support from the McGill University Doctoral Internship program through student exchange program for the SPARK Academy. The authors are grateful to McMedHacks for providing foundational information on python programming for medical image analysis as part of the 2023 SPARK Academy program. This research was funded by the Lacuna Fund for Health and Equity (PI: Udunna Anazodo, grant number 0508-S-001) and National Science and Engineering Research Council of Canada (NSERC) Discovery Launch Supplement (PI: Udunna Anazodo, grant number DGECR-2022-00136).

\newpage
\bibliographystyle{unsrt}  
\bibliography{references}

\end{document}